# A Fractal Analysis of Magnetograms within Active Regions

B. Rajkumar[1]; S. Haque[1]

**Abstract**

The magnetograms of sixteen Active Regions, observed during June – November 2015, were obtained from the Joint Science Operations Center and used to determine the fractal dimensions of various magnetic field groups. The fields were divided into strong (positive and negative) and weak (positive and negative) groupings. The area-perimeter method was used to determine the fractal dimensions for the umbral and penumbral regions of the magnetograms. The fractal dimensions were found to be $1.79 \pm 0.49$ and $1.96 \pm 0.29$ for strong and weak magnetic fields respectively. When compared to the umbral and penumbral fractal dimensions determined by (Rajkumar, Haque and Hrudey, 2017) using white light images, an inverse relationship was found, with the umbra showing lower fractal dimensions than the penumbra for the magnetic field groups compared to the intensity images. This directly implies that there is greater complexity in the penumbral magnetic field groupings than the umbral region. This has implications for models that constrain how the magnetic field groups are contained in the cooler umbral regions.

[1] Department of Physics, University of the West Indies, St. Augustine, Trinidad, W.I., Trinidad and Tobago
Email: brandon.rajkumar@sta.uwi.edu   email:shirin.haque@sta.uwi.edu

## 1. Introduction

Fractals, popularized by Mandelbrot (1983), have long been used to understand the apparent chaotic behavior of nature by quantifying its complexity, where Euclidean geometry is not easily suitable (Voss, 1988). They refer to a special type of geometry that describes complex and self-similar phenomena which are abundant in nature and demonstrate non-linearity. Self-similarity refers to the invariance under different scales and gives rise to the measure of the fractal dimension (Bunde and Havlin, 2013). From large-scale fractal structures in laboratory turbulence, the ocean and clustering of galaxies (Bershadskii, 1990) to the mechanisms behind tumor growth (Baish and Jain, 2000), fractals serve as a useful tool in understanding complex phenomena. The complexity of these phenomena can be expressed as a real, non-integer number and is referred to as the fractal dimension or Hausdorff-Besicovitch dimension (Mandelbrot, 1983). In Euclidean geometry, a point has a dimension of zero, a line a dimension of one, and a plane a dimension of 2, while a dimension of three refers to a cube in space. For a figure on a plane, if the fractal dimension is less than 2, we have fractal structures which are scale-free, demonstrating incomplete filling of the figure under examination. The greater the difference of the fractal dimension from its Euclidean value of two, this indicates that the finer the structure is exhibited. If the fractal dimension is less than 1 in a two-dimensional fractal scenario, the structures are considered to demonstrate a scale-free hierarchy and is known as fractal dust (Georgoulis, 2012; Schroeder, 2009). It therefore can be seen that the fractal dimension is a good indicator of complexity and a measure of departure from the Euclidean geometry as is frequently encountered in nature.

Given the suitability of fractals in describing many non-linear geometry in nature, it is therefore a useful tool in its application to the geometry of sunspots and active regions (Deng, 2016; Golovko and Salakhutdinova, 2012) The size distribution of active regions has been shown to demonstrate a power law which implies fractal structures (Harvey and Zwaan, 1993).

Many features observed on the sun are driven by solar magnetic fields (Hathaway and Upton, 2014). Faculae, prominences, coronal mass ejections, solar flares, and sunspots are all due to the magnetic fields on the sun. Sunspots are one of the more prominent surface features associated with the solar magnetic field since it is the most important quantity when determining the properties of sunspots (Solanki, 2003). Georgoulis (2012) notes that fractals are excellent tools to understand the origin and nature of solar magnetism.

Sunspots are dark features on the sun's surface caused by the hindrance of upward convective flow due to the emergence of magnetic flux tubes through the photosphere. They are usually characterized by a dark central region called an umbra surrounded by a lighter feature called a penumbra. Smaller regions which contain an umbra without a penumbra are called pores. Sunspots grow and decay depending on the amount of magnetic flux rising through the surface. They can coalesce or break apart and last from a few hours to several days. All sunspots and pores are found within active regions (*ARs*). *ARs* refer to the area influenced by solar magnetic fields and are formed similarly to sunspots. They are usually bipolar and may contain single or multiple sunspots, pores, and in some cases a combination of both (Solanki, 2003).

While sunspots can be seen in white light images, special magnetogram images mapping the magnetic field strengths are needed to see *ARs*.

This complex and dynamic behavior of sunspots, and by extension *ARs*, make them the ideal phenomenon for fractal analysis. Fractal dimensions have been used extensively to understand the behavior of sunspots and *ARs*. Zelenyi and Milovanov (1991) proposed a fractal model of sunspots where fractal clusters were formed from the collection of magnetic force tubes while Chumak and Chumak (1996) determined that the fractal dimension for sunspot umbrae corresponded to turbulent structures which correspond to elastic skeletons. Rajkumar, Haque and Hrudey (2017) also determined the fractal dimensions for sunspot umbrae along with the penumbrae as $2.09 \pm 0.42$ and $1.72 \pm 0.40$ respectively. They suggested that the compressible flow may be linked to the morphology of sunspots. They also observed a positive correlation between the umbral and penumbral fractal dimensions which suggests that changes in complexity between the features may be linked.

Fractals have also been used to analyze magnetograms to understand the complexity and model the behavior of solar magnetic fields and other solar phenomena. Meunier (1999) performed a fractal analysis to determine the fractal dimensions of *ARs* where it was determined that the fractal dimensions vary with magnetic threshold and that structures of moderate sizes are more complex at lower magnetic thresholds whereas large structures exhibit less complexity. Abramenko (2005) studied magnetograms and found that multifractality is a good indicator of flaring *ARs* in contrast to the determination by Georgoulis (2012) who concluded that while fractals and multifractals were useful tools for understanding the solar magnetism, they were not suitable in predicting major solar flares in *ARs*. It was noted that an increase in multifractality is a signal that a magnetic structure is being driven to a critical state. The multifractal nature of active region magnetic fields was also confirmed by Chumak (2005). By analyzing the fractal dimensions for a solar X-ray flux time series, Chumak (2005) was able to observe great variations in the fractal dimensions, even for two consecutive days which the author concluded that fractal dimensions can be used as a good X-Ray index.

A study on the fractal dimensions of *ARs* using magnetograms conducted by Ioshpa *et al.* (2005) indicated the existence of turbulent currents at the umbra-penumbra and penumbra-photosphere boundaries. Ioshpa, Obridko and Rudenchik (2008) later showed that the magnetic field distribution is more regular in *ARs*.

Our study focusses on the comparison of active region structures (umbra and penumbra) to active region magnetic fields using fractal dimensions. In section 2, the data set used is described with emphasis on how the images were prepared and measured. In section 3, the results and analysis of the data are reported. In section 4, we discuss our findings followed by the conclusion in section 5.

## 2. Method

Our study is divided into two main sections. In the first section the magnetic fields of the magnetograms are separated into various magnetic field groups based on their magnetic field strength and polarity. The fractal dimension of each field magnetic field group is determined. In the second section, flattened intensity images are used to determine the outline of the umbra and penumbra. The umbra and penumbra outlines are then overlaid unto the magnetograms to determine which magnetic field groups are present within the umbral and penumbral regions of *ARs*.

The data set used in this study is the same data set used by Rajkumar, Haque and Hrudey (2017) which contains 16 *ARs* observed between 16th June 2015 and 3rd November 2015. The 4k *HMI* full disk Color Magnetograms and 4k *HMI* full disk Flattened Intensity images which correspond to the data set were obtained from the Joint Science Operations Center (*JSOC*) *HMI* Archive. We show one typical example of the data treatment for our sample of 16 *ARs*, for AR 12367 on 16th June 2015 as seen in Figures 1 and 2. The *ARs* used by Rajkumar, Haque and Hrudey (2017) were then isolated and cropped from both magnetogram and flattened intensity full disk images as seen in Figures 1 and 2 respectively.

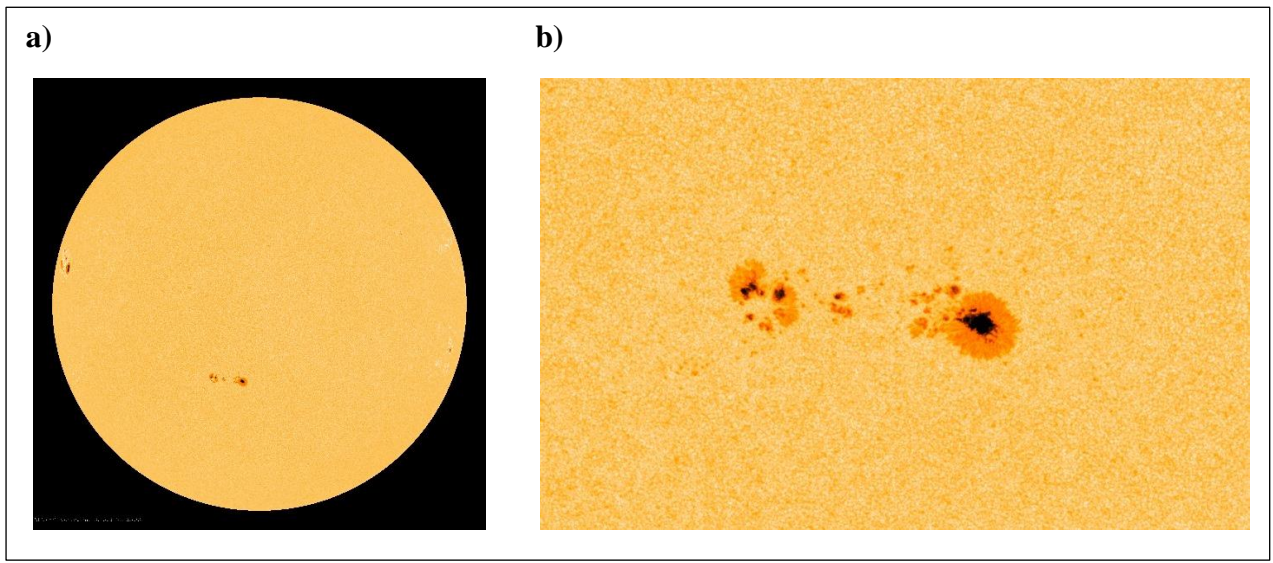

**Figure 1: a)** Flattened intensity full disk image and **b)** Flattened intensity cropped image containing AR 12367 on 16th June 2015

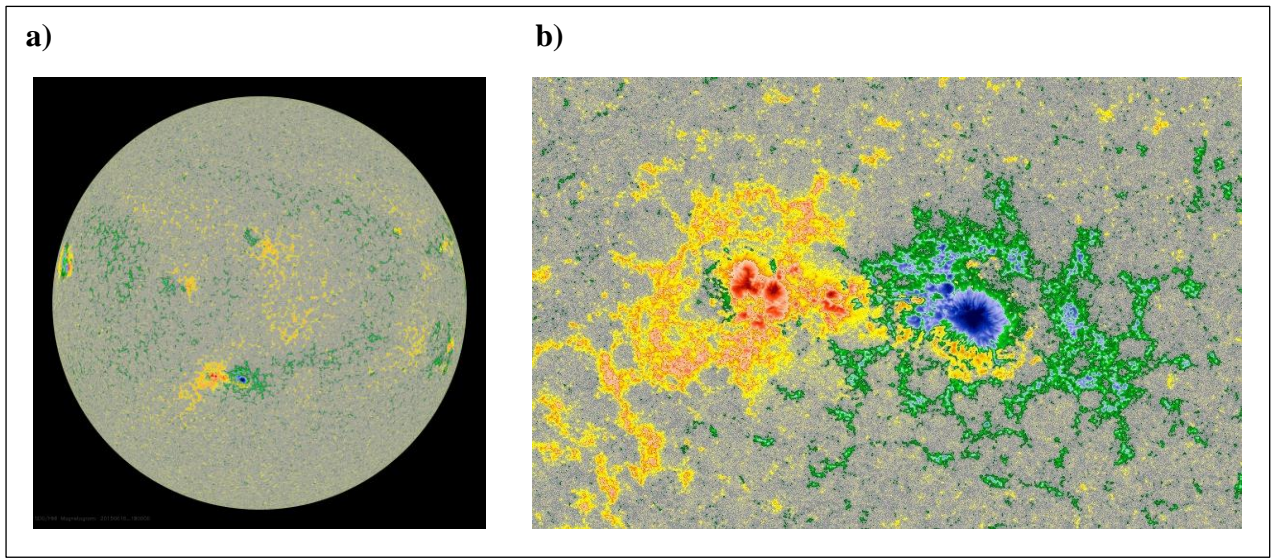

**Figure 2: a)** Color magnetogram full disk image and **b)** color magnetogram cropped image containing AR 12367 on 16[th] June 2015

The *HMI* Magnetic Field Color Table (Hoeksema, 2014) which assigns the following magnetic thresholds was used to separate the magnetic fields of the *ARs* into four main threshold groups, hereafter referred to magnetic field groups: -

- Weak positive fields (24 to 236 G) – green
- Weak negative fields (-236 to -24 G) – yellow/orange
- Strong positive fields (237 to 1500 G) – blue
- Strong negative fields (-1500 to -237 G) – red

Strong and weak fields independent of polarity were also considered and separated as follows: -

- Weak fields (-236 to -24 G) and 24 to 236 G) – yellow/orange and green
- Strong fields (-1500 to -237 G and 237 to 1500 G) – red and blue

The analysis requires the determination of areas and perimeters of the umbral and penumbral regions of the magnetograms. This was done by use of the tool ImageJ. This is a Java-based image processing program developed at the National Institutes of Health and the Laboratory for Optical and Computational Instrumentation. It is an open source image processing program designed for scientific multidimensional images which allows the determination of sizes, perimeters and areas of complicated structures. It is very useful and suitable to determine the perimeters and areas for sunspots and *ARs* as for this study.

The color threshold tool of ImageJ was used to outline each magnetic field group as seen in the example shown in Figures 3a and 3b. The measure tool was then used to determine the areas and perimeters of each magnetic field group. Areas are reported in squared pixels while the perimeters are reported in pixels. These measurements are used to determine the fractal dimensions of each magnetic field group.

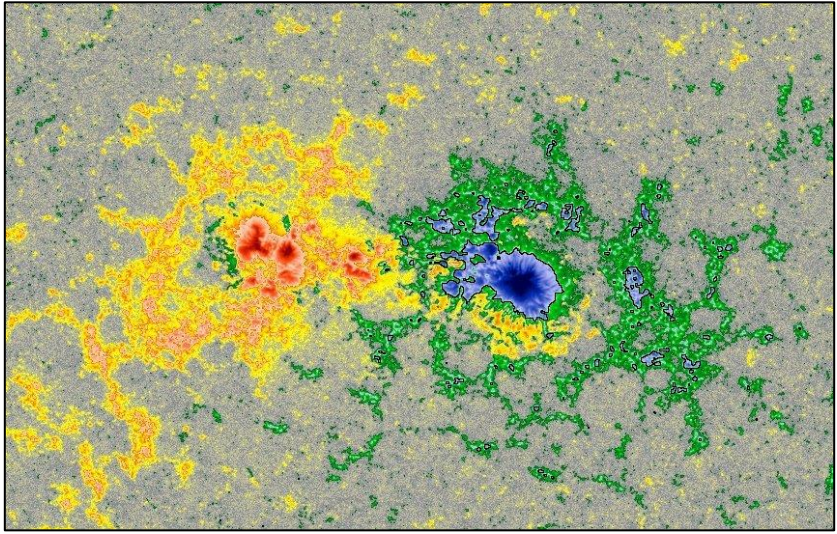

**Figure 3a)** Cropped color magnetogram of AR 12367 on 16[th] June 2015 with selected strong positive magnetic fields outlined in black

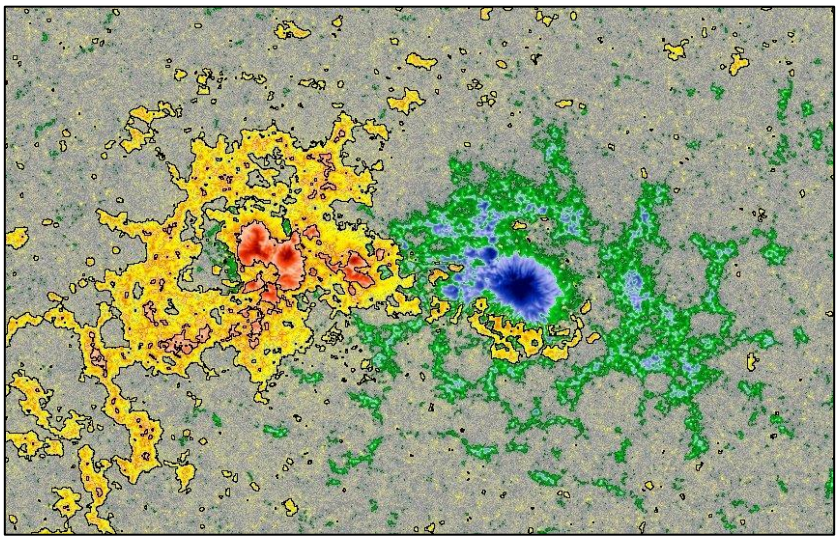

**Figure 3b)** Cropped color magnetogram of AR 12367 on 16[th] June 2015 with selected weak negative fields outlined in black.

The fractal dimensions were determined using the same area-perimeter relation utilized by (Chumak and Chumak, 1996; McAteer, Gallagher and Ireland, 2005; Meunier, 1999), Rajkumar, Haque and Hrudey (2017). This method is also recommended by Takayasu (1990) and Feder (2013) as it is very useful because of its sensitivity to the complexity of structures

(Meunier, 1999).

The area of the magnetic field group, *S*, is related to the perimeter of the magnetic field group, *L*, by

$$S \sim L^q.$$

Therefore,

$$q = \frac{\log S}{\log L},$$

where *q* is related to the fractal dimension, *d*, by

$$d = \frac{2}{q}.$$

If the structure of the magnetic field groups were uniform or circular *d* would be close to 1, however, as the structure of the magnetic fields become more complex and less uniform, *d* approaches 2 (Mandelbrot, 1983) As such, the fractal dimension, *d*, can be regarded as synonymous with complexity as noted by Meunier (1999) in the study of the fractal analysis of Michelson Doppler Imager (MDI) magnetograms.

The flattened intensity images of the *ARs* used in the previous study (Rajkumar, Haque and Hrudey, 2017) were then overlaid onto the corresponding color magnetogram. The outlines of the umbral and penumbral regions observed in the flattened intensity images were traced on to the corresponding color magnetograms. The regions on the magnetograms outlined by the umbra and penumbra of each active region were then extracted from the corresponding magnetograms (as seen in Figure 4). This allows for the determination of the magnetic fields within the umbral and penumbral regions of these *ARs*.

It should be noted that since the penumbra outline is used to determine the penumbral region, the areas and magnetic fields enclosed by the penumbra outline are considered as part of the penumbral region.

**a)**                                                                                  **b)**

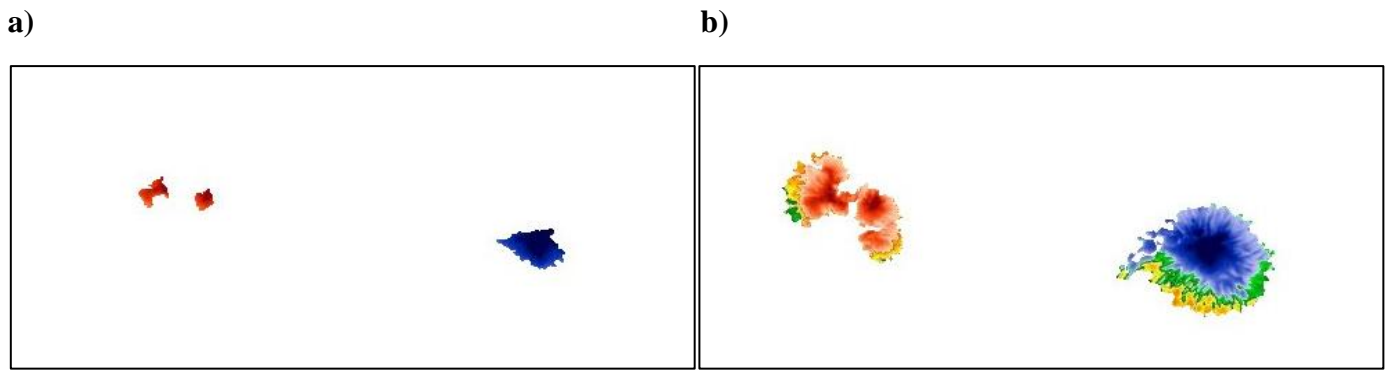

**Figure 4:** Images showing magnetic fields contained within the outline of **a)** sunspot umbra and **b)** sunspot penumbra of AR 12367 observed on 16$^{th}$ June 2015

The area of each *AR's* umbra and penumbra was then measured using ImageJ's threshold and measure tools. The areas of each magnetic field group contained within the outline of the umbra and penumbra regions were also measured (Figures 5). The areas of each magnetic field group were then compared to the areas of the umbra and penumbra separately to determine how the magnetic field groups were distributed within the umbral and penumbral regions.

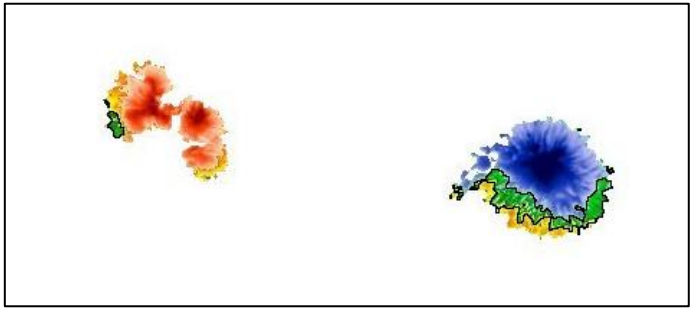

**Figure 5)** Image showing magnetic fields contained within the outline of the penumbra of AR 12367 on 16th June 2015 with selected weak positive fields outlined in black.

## 3. Results and Analysis

Figures 1 and 2 show a typical example of how the data was analyzed for the 16 ARs and their corresponding magnetograms. The *ARs* of interest were cropped from their respective HMI flattened intensity images and HMI color magnetogram images. It was ensured that cropped sections were of the same dimensions and of the exact same location to ensure accuracy in the overlay. It should also be noted that the *ARs* selected by Rajkumar, Haque and Hrudey (2017) were located near the center of the solar disk to avoid distortions caused by the Wilson effect (Loughhead and Bray, 1958) where the foreshortening is negligible.

Figures 3a and 3b show examples of how the magnetic field groups were outlined. These field groups were taken from the *HMI* Magnetic Field Color Table (Hoeksema, 2014) obtained from the *JSOC*. These outlines are analogous to the magnetic fields thresholds used by (Jurčák, et al., 2018), Meunier (1999) as well as, the magnetic field contours used by (Ioshpa, Mogilevskii, Obridko and Rudenchik, 2005), Régnier and Canfield (2006). By using the color magnetograms and the magnetic field groups outlined in the method, the magnetic fields can be compared by strength and polarity to their corresponding umbral and penumbral structures. From these outlines, the area and perimeter of the magnetic fields can be determined. Figure 6 shows the plots of log area $S$ against log perimeter $L$. The error bars included in Figure 6 represents the standard error from the regression analysis. From Figure 6, Table 1 summarizes the values of $q$ and the fractal dimension $d$ for the various magnetic field thresholds. The errors reported in Table 1 are at the 95% confidence level.

From Table 1, it is seen that the fractal dimensions for positive weak (1.77 ± 0.20) and positive strong (1.57 ± 0.37) fields are lower than the fractal dimensions for negative weak (1.83 ± 0.25) and negative strong (1.67 ± 0.42) fields respectively. Since higher fractal dimension indicate greater geometric complexity, this means that our analysis shows that negative fields create more complex structures than positive fields. It is also noted that the fractal dimensions for positive weak, negative weak and weak ($1.96 \pm 0.29$) fields are higher than the corresponding positive strong, negative strong and strong ($1.79 \pm 0.49$) fields. In both cases, the strong fields have lower fractal dimensions independent of whether it is positive or negative field strength. This indicates that a relationship between weak and strong magnetic fields exist where weak fields produce more complex structures.

Figures 4a and 4b show examples of the magnetic fields present within the umbra and penumbra respectively. This is achieved by overlaying the *HMI* flattened intensity images over the *HMI* color magnetograms and extracting the magnetic fields which are encompassed by the outlines of the umbra and penumbra. This was done for each active region in the data set used. When comparing the areas of each magnetic field group to the areas of the umbra and penumbra, it was noted that none of the umbral regions contained any weak magnetic fields, however it was possible for the umbral regions to contain only positive or only negative magnetic fields. This is evident in Figure 4a. It was also observed that penumbral regions contained a mixture of both strong and weak magnetic fields. While umbral regions could also contain only positive or only negative magnetic fields, the magnetic fields would be a mixture of weak and strong positive fields or weak and strong negative fields. This can also be seen in Figure 4b.

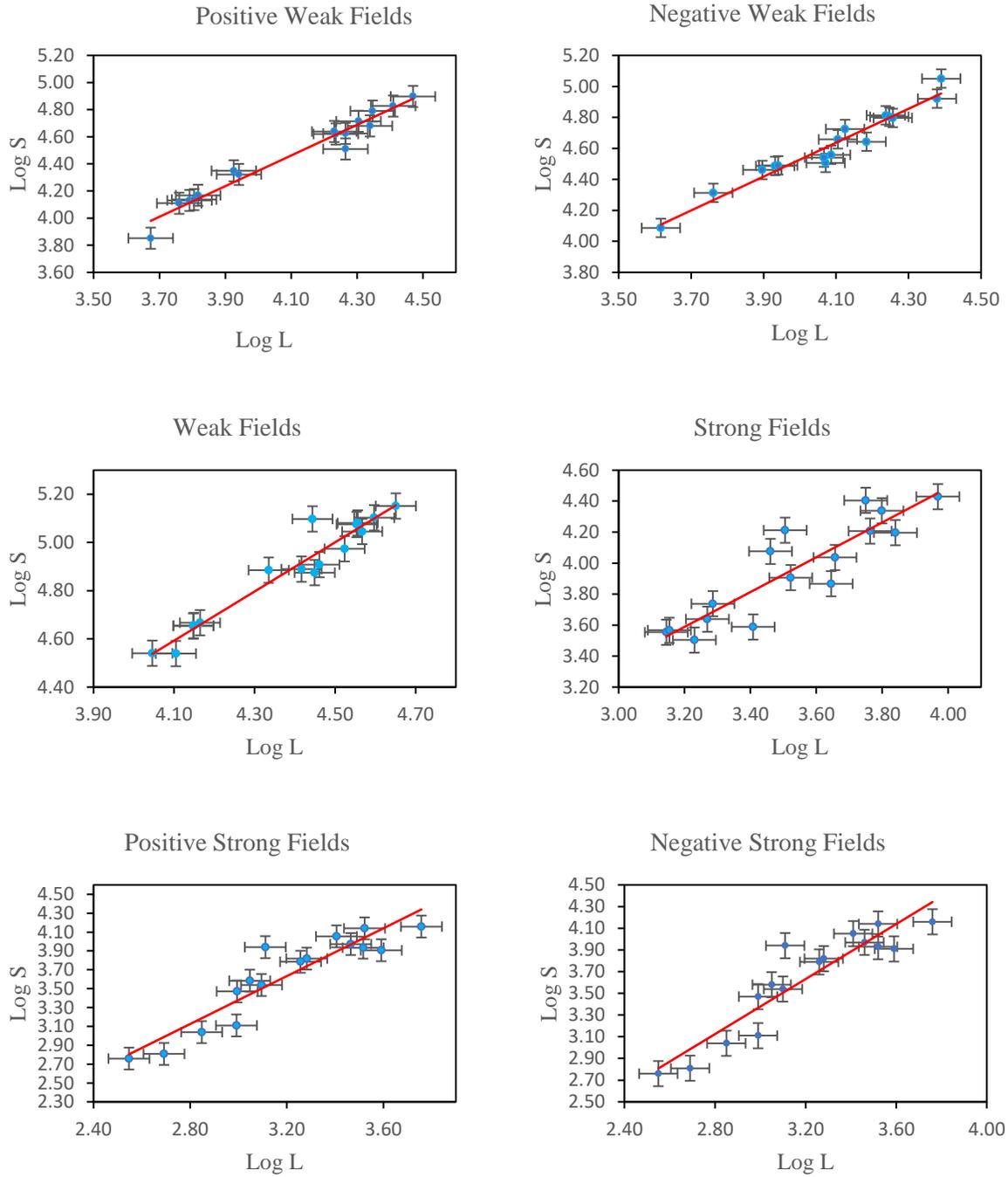

**Figure 6** Determination of the fractal dimension, d, for various magnetic field thresholds from the plots of log area *S* and log perimeter *L*.

**Table 1** Values of q, fractal dimensions, d, for various magnetic field thresholds.

| Magnetic Field Threshold | q | d |
|---|---|---|
| Positive Weak | 1.13 ± 0.13 | 1.77 ± 0.20 |
| Negative Weak | 1.09 ± 0.15 | 1.83 ± 0.25 |
| Weak | 1.02 ± 0.15 | 1.96 ± 0.29 |
| Positive Strong | 1.27 ± 0.30 | 1.57 ± 0.37 |
| Negative Strong | 1.20 ± 0.30 | 1.67 ± 0.42 |
| Strong | 1.12 ± 0.31 | 1.79 ± 0.49 |

## 4. Discussion

By measuring the areas of the various magnetic thresholds contained in both the umbra and penumbra, it was determined that strong magnetic fields dominate the *ARs*, as expected, since magnetic fields are the driving force behind active region formation (Solanki, 2003). It was observed that umbral regions contain only strong magnetic fields while penumbral regions contain both. According to Jurčák *et al.* (2018), stronger and more vertical magnetic fields are found within the umbra than within the penumbra. It is noted that the magnetic field strength decreases and becomes more horizontal with distance from the umbral core (Jurčák *et al.*, 2018). This suggests that while strong magnetic fields dominate the penumbra, the presence of weaker magnetic fields in the penumbra can account for differences in structure and behavior between the two regions. This would also account for the turbulence noted by Ioshpa *et al.* (2005). This can be understood by our finding of the respective fractal dimensions.

Rajkumar, Haque and Hrudey (2017), determined the fractal dimensions of 2.09 and 1.72 respectively for the umbra and penumbra from the sixteen *ARs* studied on the photosphere. Since the umbra has a higher fractal dimension than that of the penumbra, this indicated that the complexity of the structure within the umbra is higher than that of the penumbra. Rajkumar, Haque and Hrudey (2017) found that Weitz *et al.* (1985) study on gold colloids yielded similar fractal dimensions, which were driven by diffusion-limited and reaction-limited kinetics caused by ionization of the colloid, and that made an interesting analogy for the formation of umbrae and penumbrae in the *ARs* on the photosphere.

They proposed that this gold colloid analysis may be used as an analog model for Jaeggli (2011) model of sunspot formation where it is proposed that sunspots being cooler regions on the surface of the sun, allow for molecular hydrogen to exist in the umbra compared with ionized hydrogen in the penumbral regions with elevated temperatures. Jaeggli, Lin and Uitenbroek (2012) suggest that the formation of molecules provides a mechanism for isothermal concentration of the umbral magnetic field, and that this may explain the observed rapid increase in umbral magnetic field strength relative to temperature. This is consistent with our finding from our sample, of the magnetic field strengths in the umbra, where strong magnetic fields dominate this region exclusively, whereas there is mixed field strengths in the penumbral region.

Among other studies which determine the fractal dimension of magnetograms, we note that Meunier (1999) used the perimeter/area relation as employed in this study and found that the fractal dimension increased with the area of the *ARs* ranging from 1.48 to 1.68. Georgoulis (2012) determined the scale-free fractal dimension for magnetograms in *ARs* using the box counting method, and found it to range between 1.2 to 1.6. This is somewhat lower than the values we determined for the strong and weak magnetic fields. However, a major departure in our study is the inclusion of umbral and penumbral fractal dimension with giving consideration to strong and weak magnetic fields. This can explain the discrepancy in our values determined compared to the values noted. Our study shows that the fractal dimension for strong magnetic fields $1.79 \pm 0.49$ is lower than that of weak magnetic fields $1.96 \pm 0.29$. This indicates that the structures associated with weak magnetic fields are more complex than those of strong magnetic fields. The same trend is evident if we consider the average fractal dimension for weak and strong fields, without considering the polarity of the magnetism. The average fractal dimension for weak magnetic fields is 1.85 with the average value of the strong magnetic field being 1.68 from our studies. As suggested earlier, while strong magnetic fields dominated both umbral and penumbral regions, the presence of weak magnetic fields within penumbral regions may account for differences in the structure and behavior between the two regions. Considering the penumbral regions to be represented by the weak magnetic fields, the comparison shows an inverse complexity relationship between the structure of magnetic fields and the structure of the physical features they produce. The turbulence noted by Ioshpa *et al.* (2005) at the interface of the umbra and penumbra could be responsible for the greater complexity in the magnetogram in the penumbra as determined by our study.

Some of the earlier determinations of the fractal dimensions with magnetic thresholds yield mixed results. Ruzmaikin, Sokoloff and Tarbell (1991) found the fractal dimension decreasing from 1.71 to 1.49 between 200 and 800 G. Between 12 and 48 G, the variation went from 1.42 to 1.19. Balke *et al.* (1993) obtained an average fractal dimension of 1.54 for the magnetic threshold between 200 and 900 G. Another study by Nesme-Ribes, Meunier and Collin (1996) found the fractal dimension for 1000 faculae in *ARs* to vary between 1.60 and 1.72 based on the method utilized. Meunier (1999) notes that these discrepancies may be explained by different samples and spatial resolutions which will be applicable to our study as well.

Our study further supports Jaeggli (2011) and Jaeggli, Lin and Uitenbroek (2012) model which shows that molecular hydrogen formed within the cooler central regions of the umbra become ionized due to the increase in temperature as they move towards the umbra-penumbra boundary. The increase in ionized hydrogen near the boundary creates a high outer pressure which traps strong magnetic fields within the central region, preventing the active region from decaying more rapidly (Jaeggli, 2011). The survey and analysis reported in Jaeggli, Lin and Uitenbroek (2012) provide observational evidence that significant $H_2$ molecule formation is present in sunspots that are able to maintain maximum fields greater than 2500 G. Due to the entrapment of the magnetic fields, we suggest that the structure in the magnetogram umbra will be tighter bound and enclosed which will allow for less complexity in geometrical structure as we have found.

The lower fractal dimension for strong magnetic fields within the umbra corresponds to the trapped magnetic fields explained by Jaeggli (2011) and Jaeggli, Lin and Uitenbroek (2012). The reduced freedom of motion leads to a less complex magnetic structure while the higher fractal dimension for weak magnetic fields, within the penumbra, can be attributed to the unbound magnetic fields within that region and turbulence therein, literally causing leakage, loose structure formation and therefore generating greater complexity. It is very interesting that the model discussed matches the totally independent geometrical measure found in our study.

## 5. Conclusion

The fractal dimension of strong magnetic fields was found to be $1.79 \pm 0.49$ while weak magnetic fields had a higher fractal dimension of $1.96 \pm 0.29$ consistent in trend with the average fractal dimension values 1.85 and 1.68 respectively. It was also determined that the umbra contained only strong magnetic fields while the penumbra contained both strong and weak fields. It is suggested that the presence of weak fields in the penumbra account for the variation of structure between the umbra and penumbra and the increase in complexity thereby in the penumbral regions. When compared to the fractal dimensions of the umbra and penumbra of the flattened intensity *ARs* determined by Rajkumar, Haque and Hrudey (2017), an inverse complexity relationship was found, where the less complex structure of the strong magnetic fields contribute to the more complex physical structure of the umbra, while the less complex structure of the penumbra is associated with the more complex structure of weak magnetic fields. This inverse complexity relationship also supports Jaeggli (2011) and (Jaeggli, Lin and Uitenbroek, 2012) model of sunspot formation.

**Acknowledgments:** The authors thank the referees whose comments were very valuable in improving the manuscript.

**Disclosure of Potential Conflicts of Interest**

The authors declare they have no conflicts of interest.

Fractal Analysis of Magnetograms				B. Rajkumar & S. Haque# References

Abramenko, V. I.: 2005, Multifractal analysis of solar magnetograms, *Solar Phys*. **228**, 29 doi:10.1007/s11207-005-3525-9.

Baish, J. W. and R. K. Jain: 2000, Fractals and cancer, *Cancer research*. **60**, 3683.

Balke, A., C. Schrijver, C. Zwaan and T. Tarbell: 1993, Percolation theory and the geometry of photospheric magnetic flux concentrations, *Solar Phys*. **143**, 215 doi:10.1007/BF00646483.

Bershadskii, A. G.: 1990, Large-scale fractal structure in laboratory turbulence, astrophysics, and the ocean, *Soviet Physics Uspekhi*. **33**, 1073 doi: 10.1070/PU1990v033n12ABEH002669.

Bunde, A. and S. Havlin: 2013. Fractals in science, Springer.

Chumak, O. and Z. Chumak: 1996, Sunspots. The model of "elastic sceletons". Estimation of sunspot umbra fractal dimension, *Astronomical and Astrophysical Transactions*. **10**, 329.

Chumak, O.: 2005, Self-similar and self-affine structures in the observational data on solar activity, *Astronomical & Astrophysical Transactions*. **24**, 93 doi:10.1080/10556790500126472.

Chumak, O. V. C. Z. N.: 1996, Sunspots. The Model of "Elastic Sceletons". Estimation of Sunspot Umbra Fractal Dimension, *The Journal of the Eurasian Astronomical Society*. **10**, 329 doi: 10.1080/10556799608205449.

Deng, L.: Year, Multi-fractal Property and Long-Range Correlation of Chaotic Time Series, IEEE, 1361 doi: 10.1109/ICISCE.2016.290.

Feder, J.: 2013. Fractals, Springer Science & Business Media, doi: 10.1007/978-1-4899-2124-6.

Georgoulis, M. K.: 2012, Are solar active regions with major flares more fractal, multifractal, or turbulent than others?, *Solar Phys*. **276**, 161 doi: 10.1007/s11207-010-9705-2.

Golovko, A. and I. Salakhutdinova: 2012, Fractal properties of active regions, *Astronomy reports*. **56**, 410 doi: 10.1134/S1063772912050034.

Harvey, K. L. and C. Zwaan: 1993, Properties and emergence patterns of bipolar active regions, *Solar Phys*. **148**, 85 doi: 10.1007/BF00675537.

Hathaway, D. H. and L. Upton: 2014, The solar meridional circulation and sunspot cycle variability, *Journal of Geophysical Research (Space Physics)*. **119**, 3316 doi:10.1002/2013JA019432.

Hoeksema, T.: 2014, in HMI_M.ColorTable (ed.), pdf, http://jsoc.stanford.edu, http://jsoc.stanford.edu/jsocwiki/MagneticField.

Ioshpa, B., E. Mogilevskii, V. Obridko and E. Rudenchik: Year, Some Fractal Properties of Solar Magnetic Fields.

Ioshpa, B., V. Obridko and E. Rudenchik: 2008, Fractal properties of solar magnetic fields, *Astronomy Letters*. **34**, 210 doi: 10.1134/S1063773708030080.

Jaeggli, S. A.: 2011, An observational study of the formation and evolution of sunspots. doi:10.1088/0004-637X/745/2/133.

Jaeggli, S. A., H. Lin and H. Uitenbroek: 2012, On molecular hydrogen formation and the magnetohydrostatic equilibrium of sunspots, *The Astrophysical Journal*. **745**, 133 doi:10.1088/0004-637X/745/2/133.

Jurčák, J., R. Rezaei, N. B. González, R. Schlichenmaier and J. Vomlel: 2018, The magnetic nature of umbra–penumbra boundary in sunspots, *Astron. Astrophys*. **611**, Article number doi: 10.1051/0004-6361/201732528.

Loughhead, R. and R. Bray: 1958, The Wilson effect in sunspots, *Australian Journal of Physics*.
14